\documentclass[11pt]{article}
\usepackage{multicol}                       	
\usepackage{eurosym}                       	
\usepackage[noblocks]{authblk}		    %
\usepackage{tabularx}                        	%
\usepackage{graphicx}                        %
\usepackage{dcolumn}                        %
\usepackage{bm}                        			%
\usepackage{amsmath}                       %
\usepackage{amssymb}                      %
\usepackage[squaren,Gray]{SIunits}  	%
\usepackage{float} 								
\usepackage{units}								%
\usepackage{amsmath}						%
\usepackage{amssymb}						%
\usepackage{color}								%
\usepackage[dvipsnames]{xcolor}		%
\usepackage{hyperref}						%
\usepackage{url}									%
\usepackage[left,columnwise]{lineno}	
\usepackage{comment}

\newcommand{\blue}{\textcolor{black}}

\makeatletter
\def\@maketitle{%
  \newpage
  \null
  \vskip 0.em%
  \vspace{-1.cm}
  \begin{center}%
  \let \footnote \thanks
    {\huge \bfseries \@title \par}%
    \vskip .0em%
    {\normalsize
      \lineskip 5.em%
      \begin{tabular}[t]{c}%
        \@author
      \end{tabular}\par}%
  \end{center}%
  \par
  \vskip .em}
\makeatother

\newcommand{\APC}{\small Universit\'{e} de Paris Cit\'{e}, CNRS, APC, Paris, France}
\newcommand{\Bordeaux}{\small Universit\'e de Bordeaux, CNRS, LP2I Bordeaux, Gradignan, France}
\newcommand{\IJC}{\small Universit\'e Paris-Saclay, CNRS/IN2P3, IJCLab, Orsay, France}
\newcommand{\LNCA}{\small LNCA Underground Laboratory, CNRS, EDF Chooz Nuclear Reactor, Chooz, France}

\newcommand{\PadUNI}{\small Dipartimento di Fisica e Astronomia dell’Universit\`{a} di Padova and INFN Sezione di Padova, Padova, Italy}
\newcommand{\PadINFN}{\small INFN, Sezione di Padova, Padova, Italy}
\newcommand{\PUC}{\small Department of Physics,  Pontif\'icia Universidade Cat\'olica do Rio de Janeiro, Rio de Janeiro, Brazil}
\newcommand{\SUBA}{\small Nantes Université, IMT-Atlantique, CNRS/IN2P3, Subatech, Nantes, France}
\newcommand{\SYSU}{\small Sun Yat-sen University, Guangzhou, China}
\newcommand{\UCI}{\small Department of Physics and Astronomy, University of California at Irvine, Irvine, CA, USA}
\newcommand{\ILANCE}{\small ILANCE, CNRS - University of Tokyo International Research Laboratory, Kashiwa, Chiba, Japan}

\newcommand{\LSDa}{LSD}

\newcommand{\PCa}{PCR}

\newcommand{\PIa}{PIR}

\newcommand{\MC}{multi-calorimetry}
\newcommand{\SC}{single-calorimetry}

\newcommand{\DC}{dual-calorimetry}

\begin{document}

\author[1,2,3]{Anatael~Cabrera\thanks{\scriptsize Contact:~\texttt{anatael@in2p3.fr}}}
\author[1,2,4]{Yang~Han\thanks{\scriptsize  Contact:~\texttt{hany88@mail.sysu.edu.cn}}}
\author[5]{Steven Calvez}
\author[6]{Emmanuel~Chauveau}
\author[1]{Hanyi~Chen}
\author[2]{Hervé~de Kerret\thanks{\scriptsize Deceased}}
\author[7]{Stefano~Dusini}
\author[2,8]{Marco~Grassi}
\author[5]{Leonard Imbert} 
\author[4]{Jiajun~Li}
\author[9]{Roberto~Carlos~Mandujano}
\author[1]{Diana~Navas-Nicolás}
\author[1,10]{Hiroshi~Nunokawa}
\author[2]{Michel~Obolensky}
\author[9]{Juan~Pedro~Ochoa-Ricoux}
\author[5,11]{Guillaume~Pronost}
\author[5]{Benoit Viaud}
\author[5]{Frédéric~Yermia}

\affil[1]{\IJC}
\affil[2]{\APC}
\affil[3]{\LNCA}
\affil[4]{\SYSU}
\affil[5]{\SUBA}
\affil[6]{\Bordeaux}
\affil[7]{\PadINFN}
\affil[8]{\PadUNI}
\affil[9]{\UCI}
\affil[10]{\PUC}
\affil[11]{\ILANCE}

\title{ \parbox{1.1\textwidth} {\bf Multi-Calorimetry in Light-based Neutrino Detectors}}
\maketitle

\noindent
\begin{center}
	\parbox{0.94\textwidth}
	{\small \bf \noindent
 Neutrino detectors are among the largest photon detection instruments, built to capture scarce photons upon energy deposition.
Many discoveries in neutrino physics, including the neutrino itself, are inseparable from the advances in photon detection technology, particularly in photo-sensors and readout electronics, to yield ever higher precision and richer detection information. 
The measurement of the energy of neutrinos, referred to as {\it calorimetry}, can be achieved in two distinct approaches: photon-counting, where single-photon can be counted digitally, and photon-integration, where multi-photons are aggregated and estimated via analogue signals.
The energy is pursued today to reach permille level systematics control precision in ever-vast volumes, exemplified by experiments like JUNO.
The unprecedented precision brings to the foreground the systematics due to calorimetric response entanglements in energy, position and time that were negligible in the past, thus driving further innovation in calorimetry.
This publication describes a novel articulation that detectors can be endowed with multiple photon detection systems.
This {\it multi-calorimetry} approach opens the notion of {\it dual-calorimetry} detector, consisting of both photon-counting and photon-integration systems, as a cost-effective evolution from the {\it single-calorimetry} setups used over several decades for most experiments so far.
The \DC\ design exploits unique response synergies between photon-counting and photon-integration systems, including correlations and cancellations in calorimetric responses, to maximise the mitigation of response entanglements, thereby yielding permille-level high-precision calorimetry.
}
\end{center}

\begin{multicols*}{2}


Neutrino detectors are amongst the most gigantic photonics systems.
Photons, produced via rare neutrino interactions in light-based detectors, travel quasi-unimpededly over long distances across transparent monolithic media, and then, they can be detected by photo-sensors.
This is the basis of one of the most successful workhorse technologies, leading to many neutrino-related discoveries over several decades, as evidenced by several Nobel prizes~\cite{Nobel1995, Nobel2002, Nobel2015}.
The art of neutrino detection is tightly linked to advances in photon detection, and it is expected to remain so for the foreseeable future.

The main challenge of neutrino detectors stems from their monumental size, needed to compensate for the tiny neutrino detection probability, so the detection of photons takes place over distances of up to a hundred meters across.
Most neutrino detection is reduced to light detection with photonic instruments, i.e., photo-sensor and readout electronics; meanwhile, the detection systematics control is crucial to achieving high-precision measurement.
However, one of the most challenging applications of these instruments is that the specific photon detection solution remains primarily constrained by the detector dimension and corresponding cost-effective compromises while pursuing high precision.
A novel light detection technique, called {\it dual-calorimetry}, is explored and tailored to excel within these constraints, which is the main subject of this publication.

\begin{figure*}[t!]
	\vglue 0.2cm
	\centering
	\includegraphics[scale=0.23]{Figure_1.pdf}
        \caption{
	{\bf(a.) Light-based Neutrino LSD Typical Configuration.}
        A neutrino detector is typically located underground, equipped with shielding systems for background reduction.
        The detection system consists of a monolithic detection medium (purple), e.g. liquid scintillator, surrounded by photo-sensors (orange) whose signals (blue) are processed through readout electronics.
	\label{Fig1.a}
	{\bf (b.) Photon Detection Systems and Dual Calorimetry Configuration.}
	A single photo-sensor (SPS) with the large surface can receive multiple photons, thus requiring its electronics to sample the signal pulse waveform and integrate it over time, i.e. photon integration.   	
	As long as the SPS is small enough such that single-photon signals dominate, it can be sufficient for its electronics to 
keep minimal functionality to discriminate single-photon from noise, i.e. photon-counting.
	The \DC\ enables both photon-integration and photon-counting calorimeters in a single \LSDa\ by having two sets of photo-sensors around the common detection volume.
        }
        \renewcommand{\thefigure}{\arabic{figure}.(b.)} 
        \label{Fig1.b}
\end{figure*}

Within today's neutrino detectors, the pioneering {\it liquid scintillator detector} (\LSDa) has been the most widely used.
Scintillating materials as the detector medium enable the detection of additional light~\cite{Birks} relative to the irreducible Cherenkov radiation~\cite{Cherenkov1934, Cherenkov1937}.
Hence, with about one hundred times more light, higher energy resolution of several percent in the MeV scale and lower energy detection threshold, e.g. tens of keV, are possible in \LSDa ~\cite{JUNO2021}.
A general view of the monolithic light-based neutrino LSD is described in Figure~\ref{Fig1.a}\blue{.(a.)}.

The \LSDa\ technology has led to major fundamental neutrino discoveries and measurements.
Those achievements include the neutrino discovery~\cite{Cowan1956}, 
the first reactor neutrino oscillation studies by the Chooz~\cite{CHOOZ1999}, Palo Verde~\cite{PaloVerde1999} and KamLAND~\cite{KamLAND2002}, 
the first accelerator-based neutrino oscillation studies by LSND~\cite{LSND1996} and KARMEN~\cite{KARMEN2002},
the discovery of the Earth's radioactivity geoneutrino by KamLAND~\cite{KamLANDgeo}, 
the first observation of the model-predicted third-neutrino-oscillation by Daya Bay~\cite{DayaBay2012}, Double Chooz~\cite{DoubleChooz2011} and RENO~\cite{RENO2012}, 
and the observation of all solar neutrino components by Borexino~\cite{BorexinoBe7, BorexinoB8, Borexinopep, Borexinopp, BorexinoCNO}.
Most recently, even the world-leading precision in neutrino-less double-beta decay searches benefits from these detectors, as demonstrated by KamLAND-zen~\cite{KamLANDZen2022}, and soon also SNO+~\cite{SNOplus2021}.

Detectors based only on Cherenkov radiation hold a similar photon detection system as \LSDa.
Larger volumes are possible for water Cherenkov detectors by the absence of scintillators due to the higher transparency in water but at the expense of scarcer light for detection. 
The discovery of neutrino oscillations led by Super-Kamiokande~\cite{SuperK1998}, the largest-built detector so far ($\sim$5$\times$10$^4$\,m$^3$), and SNO~\cite{SNO2002},
as well as the first observation of neutrinos from a supernova core-collapse explosion by the Kamiokande~\cite{Kamiokande1987} and IMB~\cite{IMB1987} experiments, rejoin this photonics technology {\it palmar\`es}.
Scaling up the same rationale, Hyper-Kamiokande~\cite{HyperK2018} detector, under construction, is reaching the unprecedented instrumented volume of $\sim$2.5$\times$10$^5$\,m$^3$.
While using the natural detection medium, there are also {\it neutrino telescopes} with a total volume of order $\sim$10$^9$\,m$^3$ \cite{IceCubeDesign, KM3Net2016, Baikal}.

The next generation neutrino \LSDa\ represented by the JUNO experiment~\cite{JUNO_YB, JUNO2021}, under construction, will reach the cutting-edge technology for the highest light detection precision with a huge detector size ($\sim$2$\times$10$^4$\,m$^3$).
Its physics target imposes the calorimetry design to reach a vertiginous control of the energy systematics at the sub-percent level total budget over one of the vastest photon detection dynamic ranges.
Indeed, the unprecedented requirement in a detector like JUNO opens a new twist in the \LSDa\ photon detection innovation, leading to the conception of the {\it dual calorimetry}, first proposed in 2014, released later on~\cite{Anatael2016, Anatael2018} and here described for the first time.

The notion of \DC\ described here aims at robust light detection with stringent systematics control via the redesign of the photon detection system.
The \DC\ relies on two synergetic and complementary photon detection systems with photon-counting and photon-integration in a single detector, as illustrated in Figure~\ref{Fig1.b}\blue{.(b.)}, and can be considered as an evolution from the {\it single calorimetry} basis, i.e. single photon detection system, used by most light-based neutrino detectors so far.
From a different perspective, the so-called \DC\ could instead exploit other physics complementary observables different from light.
Excellent examples exist when scintillation light (i.e. photons) are combined with heat, in the context of scintillating bolometers~\cite{CUPID2019}, or with the disassociated electrons in noble liquid (or gas) time projection chamber detectors~\cite{nEXO}.
However, the rest of our discussion shall focus on photonics-only \DC.

\section*{Calorimetry in \LSDa} 

The notion of {\it calorimetry} implies the measurement of {\it energy}. 
The calorimetry in \LSDa\ is achieved by measuring the light yielded upon the interaction of particles in the scintillator.
For neutrino detection, the energy estimated by light is referred to as {\it visible energy}, as opposed to the {\it neutrino energy}, which may only be inferred a posteriori.
The {\it visible energy} will be referred to as {\it energy} from now on as it is the calorimetric observable. 

However, light-derived information in monolithic transparent \LSDa\ is known to have limitations in revealing event-by-event information, such as event topology, directionality and particle identification, despite advanced analysis techniques such as pulse-shape discrimination~\cite{BorexinoPSD}.
Modifying the detector design could overcome some of those limitations.
The adoption of the opacity in \LSDa\ is expected to enable event topology from MeV energies, as pioneered by LiquidO~\cite{LiquidO2019}.
Event-by-event information may also be resolved by segmenting the \LSDa\ as done in NOvA for GeV energies~\cite{NOvA2004}, as well as by using segmented solid scintillators, e.g. MINOS~\cite{MINOS_detector}, T2K-ND280~\cite{T2Knear}, Solid~\cite{SoLid} and DANSS~\cite{DANSS}. 
Enhancing the exploitation of Cherenkov light is expected to provide event directionality as recently shown~\cite{Borexino_cherenkov,SNO2022}.
This may be further enhanced by adopting specialised formulations of liquid scintillators, such as water-based scintillators~\cite{WBLS, Yeh2011,Theia} and slow scintillators~\cite{SlowFluor, SlowLS}.
While complementary, the energy control of non-monolithic scintillator detectors can be significantly different, so its specifics are not covered below.

Sticking to the major calorimetric measurements in monolithic \LSDa, the focuses are energy scale and resolution.
The energy scale specifies the conversion relation from the detected photons to the deposited energy, and calorimetric response systematics drive its precision.
The energy resolution specifies the stochastic and non-stochastic fluctuations of detected photons given a certain energy deposition.
The inherent stochastic Poisson fluctuation is driven by the photon statistics, while systematic effects of calorimetric response cause the aggregate of all non-stochastic terms.

Regarding the calorimetric response ($R$), there are three general types of effects.
Those are
(i) {\it non-linearity} (NL),
(ii) {\it non-stability} (NS),
and (iii) {\it non-uniformity} (NU),
which are, respectively, the manifestation of deviations in the detected light output from 
a perfectly linear response to energy, 
a constant response in time, 
and a uniform response across the detection volume.
These effects are often unavoidable due to inaccuracies in their control, thus affecting the characterisation of the energy scale and resolution. 

There are known deviations from linearity in scintillator response, including excitation-quenching and Cherenkov radiation, while having understood physics reasons.
However, leading-order instrumentation effects can cause additional complex non-linearities during the inclusive photon detection processes at the stages of photo-sensors, readout electronics and reconstruction. 
Both physical and instrumental non-linearity effects lead to an effective interlaced nonlinear response, where the former originates at the generation of detectable light, and the latter can be regarded as a consequence of the conversion of detected light into readout {\it charge}.
Time-wise non-stability is induced typically by detector-wise changes in time with temperature, detection medium evolution and readout configurations.
The non-uniformity effect is the manifestation of the non-uniform light collection across the detector primarily due to solid-angle acceptance so that the collected light with the same energy becomes position-dependent. 
This is expected to be corrected using the calibration samples with known vertex positions and energies.

More challengingly, realistic scenarios arise upon the combination of NL, NS and NU effects, in which all effects are entangled in the overall measured calorimetric response.
This, in turn, may translate into response degeneracies and biasing correlations in the overall measured energy.
Worse, some of those effects may not be corrected through conventional energy calibration schemes.
This is due to limited calibration sampling of the light response across the combined linearity, time and position parameter space.
The implication of these limitations to the control of energy precision can be significant and is the motivation of the novel \DC, as will be detailed in Section~\ref{Sec4}.


\begin{figure*}[t!]
	\centering
    \vglue 0.2cm
	\includegraphics[scale=0.23]{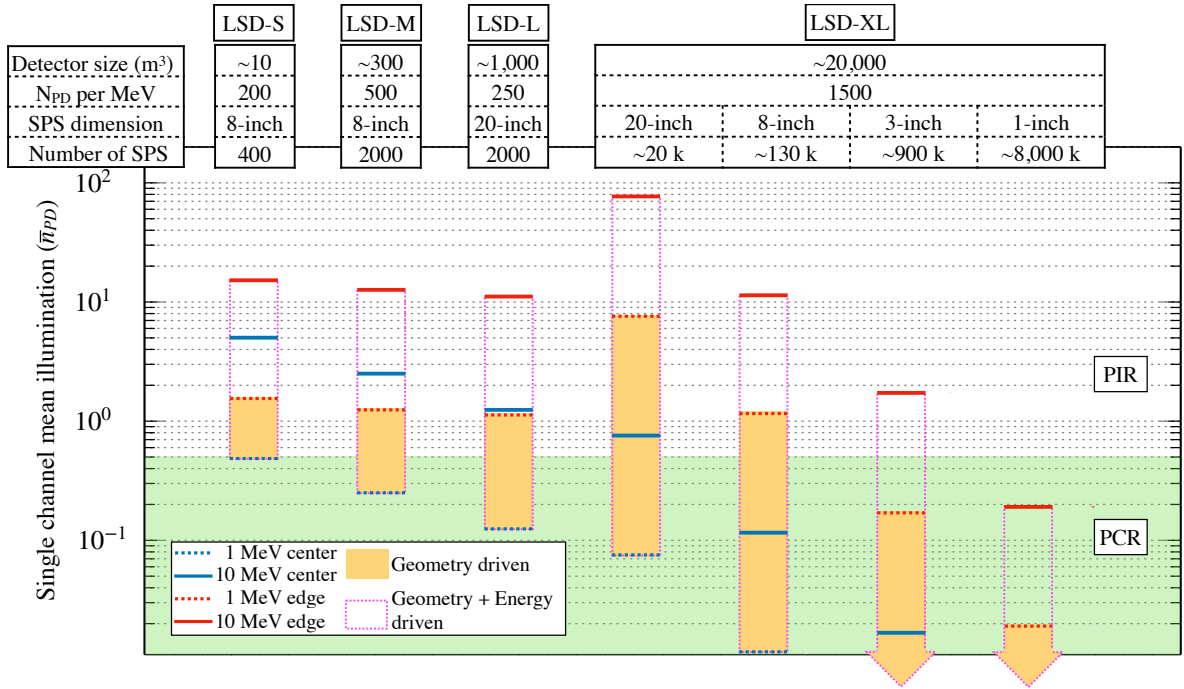}
	\caption{
	{\bf Calorimetry Light Detection Regime Illustration.}
	The light regime depends on the single channel mean illumination ($\overline{n}_{PD}$). 
	This is shown for four hypothetical detector configurations with different detector sizes: LSD-S (small), LSD-M (medium), LSD-L (large) and LSD-XL (extra-large),
	which can be considered as DoubleChooz-like, Borexino-like, KamLAND-like and JUNO-like detectors, respectively.
	{The $\overline{n}_{PD}$ is estimated to first-order by considering the detector size, spherical geometry, total photon detected ($N_{DP}$) per MeV, SPS dimension and number of SPS.
    Additional effects to $\overline{n}_{PD}$ estimation, such as imperfect spherical geometry and channel-to-channel difference are ignored for simplification purpose.}
    For each detector configuration, the $\overline{n}_{PD}$ is calculated for 
    I.) 1 MeV energy deposition at detector centre (dotted blue lines), 
    II.) 10 MeV energy deposition at detector centre (solid blue lines), 
    III.) 1 MeV energy deposition at detector edge (dotted red lines), 
    IV.) 10 MeV energy deposition at detector edge (solid red lines).
    The yellow area represents the $\overline{n}_{PD}$ range driven by detector geometry.
    The area surrounded by dotted magenta lines represents the $\overline{n}_{PD}$ range driven by both the detector geometry and the energy range.
    The $\overline{n}_{PD}$ of 0.5 is approximately considered as the bound between \PCa\ and \PIa. 
    While the notion of photon dynamic range only makes sense for $\overline{n}_{PD}$ above 0.5, it is considered to be effectively zero for $\overline{n}_{PD}$ below 0.5.
    Only LSD-XL with 3-inch or 1-inch SPS configurations is predominantly in \PCa, while the other configurations are mainly in \PIa, as in most experiments so far.
    The LSD-XL with 20-inch SPS experiences the highest photon dynamic range. 
    The number of SPS for LSD-XL detector is estimated for $\sim$80$\%$ total photosensitive coverage.
    }
	\label{Fig2}
\end{figure*}

\section*{Single Calorimetry \LSDa} 

Depending on the photon detection technology employed, the observable from a single photo-sensor (SPS), i.e. the number of photons detected per channel ($n_{pd}$), can be estimated in two distinct regimes here referred to as
{\it photon-counting regime} (\PCa)
and
{\it photon-integration regime} (\PIa).
The sum of the channel-wise $n_{pd}$ yields the total number of photons detected across the entire detector ($N_{pd}$), which is proportional to the event energy.
Single calorimetry \LSDa\ is characterised by having a single-type photon detection instrument, typically sticking to either regime. 

The \PCa\ occurs when single-photon detection is dominant.
This implies opting for a technology leading to the direct measurement, as opposed to inferred, of discrete detected single-photon.
In this context, there may be little or even no need for charge digitisation, as a simple discriminator may enable a digital (i.e. “0” and “1”) readout per detected photon; hence the photon ($n_{pd}$) dynamic range is virtually zero, as shown in Figure~\ref{Fig1.b}\blue{.(b.)}.
The calorimetry in \PCa\ features a direct $N_{pd}$ observable, immune to the charge measurement and its possible bias.
Hence, it has a minimal systematic effect, i.e. the discrimination of single-photon from noise, promisingly leading to robustness of the light detection. 

When the pile-up of multiple photons per SPS is not negligible, the detection enters into the \PIa, and light is effectively sampled and integrated as the integral charge over a certain time window, as highlighted in Figure~\ref{Fig1.b}\blue{.(b.)}.
In \PIa, the photon detection exhibits a dynamic range driven by the energy of detected events and the acceptance solid-angle of each photo-sensor.
Since $n_{pd}$ is indirectly derived, this method heavily depends on the photon-to-charge conversion, typically referred to as {\it gain} in the literature.
The inferred $n_{pd}$ is a priori sensitive to numerous systematic effects, such as photo-sensor gain non-linearity, signal pulse fluctuation and distortion, electronics noise, cross-talk among channels, charge reconstruction algorithm, etc. 
Moreover, the manifestation and impact of those effects would intensify with the increase of photon dynamic range.

Given a detector configuration, the leading order differentiator between \PCa\ and \PIa\ can be the SPS dimension, as described in Figure~\ref{Fig2}.
\PIa\ is typically mandatory when the overall light level per channel is high, causing an unavoidable pile-up.
This means that the simplest way to implement \PCa\ is to reduce the SPS detection surface so that the single-photon detection is dominant, covering variations over the entire targeted range in energy and volume.
To reach the same total photosensitive area, the \PCa\, using a small pixel, implies a larger number of independent readout channels.

More readout channels have typically led to a higher cost because, traditionally, the overall cost had decreased with maximal photocathode area per channel.
Moreover, there is a non-negligible cost of logistical overheads, such as cabling.
This remains largely true today, although dedicated solutions such as multiplexing, multi-channel readout ASICs, or even underwater readout deployment for minimising cabling enable cost reduction at reasonable system reliability. 
Traditionally, the \LSDa\ design was mainly driven by the best cost-effective option.
Consequently, \PIa\ was the choice for most experiments in the past, as illustrated in Figure~\ref{Fig2}. 

On the other hand, today's latest signal waveform digitisation technology, such as FADC-based~\cite{FADC} or analogue-memory-based~\cite{waveform_analog}, provides the ability to mitigate the systematic effects associated with the detection of multiple photons, so that a \PIa\ system may infer some of the information only native in \PCa.
Waveform solutions are, however, not inexpensive, so the cost may favour a native \PCa\ approach in suitable conditions.
Moreover, waveform-derived solutions may suffer from the impact of non-negligible systematics due to reconstruction algorithms since PIR is ultimately unavoidable with increasing pile-up.

\section*{Dual Calorimetry \LSDa}  \label{Sec4}

\subsection*{ Motivation - Calorimetry Response Correlation} 

Given either \PIa\ or \PCa\ solution, the focus is the robustness of the control of energy systematics caused by unavoidable biases effects of NL, NS and NU in calorimetry response. 
In the ideal scenario of the perfect control of calorimetry systematics, \PCa\ and \PIa\ should lead to an identical outcome.
This necessitates one of the two conditions:
a) the negligible native energy detection biases, 
or,
b) an exhaustive calibration scheme perfectly corrects all biases.
Unfortunately, most of the time, neither is true in reality.

In fact, it largely depends on the calorimetry systematics tolerance of the experiment for those biases to be effectively negligible -- or not.
While some experiments do physics within a few percent precision in the energy, the latest experiments targeting high-precision neutrino measurements are pushing the calorimetry systematics control to permille level precision~\cite{DYBcalib, DCNature}, and more notably over an ever-vaster dynamic range~\cite{JUNOcalibration}, thus bringing to the foreground some bias effects that were negligible in the past.
It is exactly in this context that the \MC\ is conceived and expected to excel.

To illustrate the impact of systematics, it is convenient to define the calorimetric response ($R$) metric as  
\begin{eqnarray}
R  & = & R_{\scriptscriptstyle o} \cdot \alpha_{\scriptscriptstyle NU} \cdot  \alpha_{\scriptscriptstyle NS} \cdot \alpha_{\scriptscriptstyle NL},
\label{ResponseSC}
\end{eqnarray}
where $R$ is characterised relative to a specific reference response $R_{\scriptscriptstyle o}$, and the normalised terms $\alpha_{\scriptscriptstyle i}$ for NU, NS and NL will absorb the response dependencies on position, time and true energy, respectively.
Subsequently, the response ($R$), expressed in arbitrary units, provides the measured energy $E = \beta \cdot R$, where $\beta$ is the absolute energy conversion constant.
Once perfectly calibrated, the effect of the $\alpha_{\scriptscriptstyle i}$ is expected to be fully corrected so that the energy scale is linear, stable, and uniform across the detector and its lifetime.

Since the NL term has two different origins, physical and instrumental, linked to {\it light} ($l$) and {\it charge} ($q$), respectively,
Equation~(\ref{ResponseSC}) may be redefined as
\begin{eqnarray}
R = R_{\scriptscriptstyle o} \cdot \alpha_{\scriptscriptstyle NU} \cdot  \alpha_{\scriptscriptstyle NS} \cdot [\alpha_{\scriptscriptstyle NL(l)} \otimes \alpha_{\scriptscriptstyle NL(q)}],
\label{ResponseSCall}
\end{eqnarray}
thus expressing, with the $\otimes$-sign, the entangled NL term as the combination of detector-wise light $\alpha_{\scriptscriptstyle NL(l)}$ and channel-wise charge $\alpha_{\scriptscriptstyle NL(q)}$ components, as if they were measured separately, which is in itself one of the challenges.

While some a priori information may exist about $\alpha_{\scriptscriptstyle NL(l)}$ through measurements with test benches~\cite{NL2015}, the most faithful characterisation is done with the detector because monolithic detectors are characterised by one key feature: the detection and detector volumes are the same.
This feature leads to an even more challenging consequence: it is difficult to directly access $\alpha_{\scriptscriptstyle NL(q)}$ with the realistic detector configuration.
And it is precisely for this reason that experiments relying on multi-detector configurations to yield inter-detector cancellation of response systematics have opted for the non-cost-effective use of several identical detectors, as originally proposed by Double Chooz~\cite{DoubleChoozproposal}, and successfully demonstrated for the entire field~\cite{DayaBaydetector, RENOexperiment}.

In the above-presented response framework, there is a key assumption: each $\alpha_i$ term is independent of one another, as suggested by the effective relation
\begin{eqnarray}
{\alpha_{\scriptscriptstyle NU} \otimes \alpha_{\scriptscriptstyle NS} \otimes \alpha_{\scriptscriptstyle NL} = 0},
\label{OrthogonalSC}
\end{eqnarray}
where the $\otimes$-sign represents qualitatively the correlation, and zero means no correlation or {\it orthogonality}.
This implies that each component can be measured and thus calibrated independently as if the response linearity, uniformity and stability bias effects were totally disjoint realms.
This abstraction is, strictly speaking, not realistic, while it can be effectively tolerable if the impact of the correlations among those effects is small.
Indeed, this has been the case in most experiments so far.

\begin{figure}[H]
    \vglue -0.1cm
	\centering
	\includegraphics[scale=0.25]{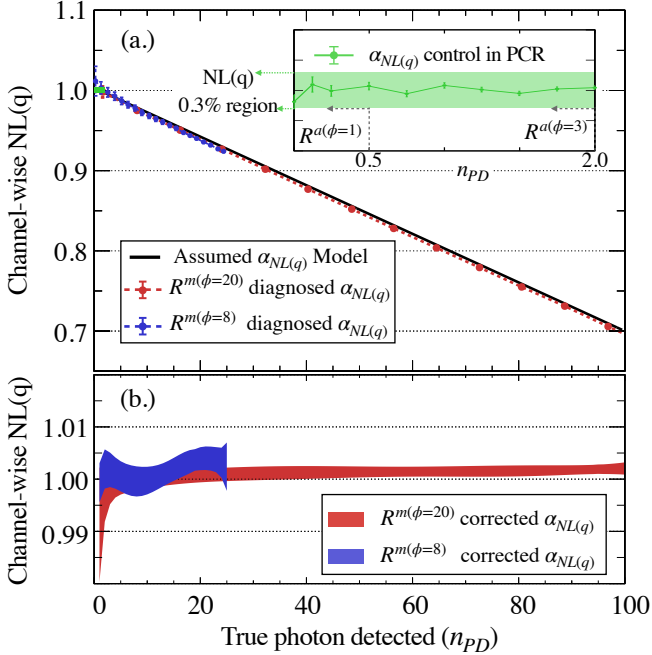}
	\caption{
	{\bf (a.) Charge Non-linearity Illustration.}
	The channel-wise $\alpha_{\scriptscriptstyle NL(q)}$ is the ratio between measured charge and true charge, here presented with the unit of $n_{PD}$.
	The LSD-XL configuration (2$\times$10$^4 m^3$) is used for illustration with a hypothetical $\alpha_{\scriptscriptstyle NL(q)}$ model indicated by the black line.
	Given the same model, the SPS responses of 20-inch ($R^{\scriptscriptstyle (\phi=20)}$) and 8-inch $R^{\scriptscriptstyle (\phi=8)}$ undergo $\sim$30$\%$ and $\sim$6$\%$ charge non-linearity within their respective photon dynamic range. 
	For the SPS responses of 3-inch ($R^{\scriptscriptstyle (\phi=3)}$) and 1-inch ($R^{\scriptscriptstyle (\phi=1)}$) with close to zero photon dynamic range, the charge linearity control can be achieved at permille level based on photon-counting~\cite{Han2020}.
	\label{Fig3.a}
	{\bf  (b.) Charge Non-linearity Control Through Dual-Calorimetry. }
	With the {\it channel-wise \DC\ calibration} approach, under $C^{m(\phi=20)}_{a(\phi=3\  or \ 1)}$ and $C^{m(\phi=8)}_{a(\phi=3 \ or \ 1)}$ configurations, the $\alpha_{\scriptscriptstyle NL(q)}$  in $R^{\scriptscriptstyle (\phi=20)}$ and $R^{\scriptscriptstyle (\phi=8)}$ can be diagnosed and then calibrated to permille level with the aid of the linear charge reference of $R^{\scriptscriptstyle (\phi=3)}$ or $R^{\scriptscriptstyle (\phi=1)}$.
    \label{Fig3.b}
	}
\end{figure}

A key element is the channel-wise photon dynamic range, where the wider it is, the harder it will be to achieve precise energy control, as illustrated in Figure~\ref{Fig3.a}\blue{.(a.)}.
So, experiments aiming for sub-percent energy precision and/or undergoing vast dynamic range are more susceptible to the complex response entanglements~\cite{DoubleChoozCalib,DYBcalib,JUNOcalibration}.

The response entanglement can be trivially illustrated with the $\alpha_{\scriptscriptstyle NL}$ and $\alpha_{\scriptscriptstyle NU}$ terms.
As $\alpha_{\scriptscriptstyle NU}$ is mainly driven by the solid angle per photo-sensor, the channel-wise detected photons $n_{pd}$ varies as $r^{-2}$, where $r$ is the distance between event position and photo-sensor.
Therefore, scanning the response in position forcibly implies the simultaneous scan of $n_{pd}$, which is directly sensitive to $\alpha_{\scriptscriptstyle NL(q)}$.
Hence, a genuine bias in linearity may manifest as an effective bias in uniformity, even if none really existed. This is illustrated with simulation as shown in Figure~\ref{Fig4}\blue{.(a.II)}.
If these were unknown, residual biases would exist after calibration.
It implies the need to modify the aforementioned response metric by adding the crossed-terms of the type $\alpha_{\scriptscriptstyle ij}$, leading to a hypothetical {\it response-tensor} representation.
Therefore, one should measure and account for all the response correlations, which typically are experimentally impractical or extremely difficult to achieve. 
The inaccurate characterisation of the unavoidable response entanglement is a main limitation for high precision energy control today.

To overcome the response entanglement challenge, there are at least two possible solutions.
One is to pursue an exhaustive calibration scheme to characterise the full response tensor, including all possible correlations.
While this may not be impossible, decades of experience reveal challenges due to limitations on calibration sources and deployments, e.g. limited available sources, positions and low calibration frequency.
These limitations would largely impact the overall precision achievable by a given calibration scheme, which has been the main effort so far and should always remain the first consideration.

The alternative is to innovate the detector design such that the impact of the inter-response correlations is essentially eliminated or significantly minimised.
As the detector's photonics system is the sole interface to the light response characterisation, the innovation is introduced here by adopting two independent photon detection systems designed to operate in \PIa\ and \PCa, respectively, inside one single detector, leading to the \DC\ design as a minimal evolution from the traditional \SC\ in \PIa.   

As detailed below, there exist synergy effects in the dual-systems, leading to the breakdown of response entanglements.
The \DC\ design aims to redefine the detector response basis where all $\alpha_{\scriptscriptstyle i}$ terms can turn effectively orthogonal in the context of stringent response control and vast dynamic range conditions.
This option can also be regarded as aiming to facilitate the calibration campaign and to spot possibly unforeseen systematics in light of unknown correlations.

While the best example today of \DC\ design is JUNO, some of the key precursory building blocks and motivation of this approach were pioneered and exercised in the energy scale definition of Double Chooz, precisely to address its \SC\ design inherent limitations~\cite{DoubleChoozCalib}.
In Double Chooz, this was effectively done by defining two \SC\ metrics, i.e. a \PIa\ and an {\it ad hoc} \PCa, from a single photon detection system, but it worked only in a very restricted low energy range at detector central area to fulfil the \PCa\ condition.
The \DC\ introduced here explores a much richer synergy effect for the entire photon dynamic range and full detection volume thanks to enabling simultaneously two independent native \PIa\ and \PCa\ photon detection systems.

\subsection*{ Dual Calorimetry formulation} 

By design, a dual calorimetry detector is endowed with a dual (two) photon detection systems, where the {\it main} system follows the generally more cost-effective \PIa\ to collect most of the light, indicated by its response as
\begin{eqnarray}
R^m & = & R^m_{\scriptscriptstyle o} \cdot \alpha^m_{\scriptscriptstyle NU} \cdot  \alpha^m_{\scriptscriptstyle NS} \cdot [\alpha^m_{\scriptscriptstyle NL(l)} \otimes \alpha^m_{\scriptscriptstyle NL(q)}],
\label{ResponseDC-M}
\end{eqnarray}
while a dedicated {\it auxiliary} system works in \PCa\ with its response as
\begin{eqnarray}
R^a & = & R^a_{\scriptscriptstyle o} \cdot \alpha^a_{\scriptscriptstyle NU} \cdot  \alpha^a_{\scriptscriptstyle NS} \cdot [\alpha^a_{\scriptscriptstyle NL(l)} \otimes \alpha^a_{\scriptscriptstyle NL(q)}].
\label{ResponseDC-A}
\end{eqnarray}
Thanks to the effectively zero photon dynamic range of \PCa\ thus charge immunity, it turns out $\alpha^a_{\scriptscriptstyle NL(q)} \to 1$ in $R^a$ and it becomes
\begin{eqnarray}
R^a & \approx & R^a_{\scriptscriptstyle o} \cdot \alpha^a_{\scriptscriptstyle NU} \cdot  \alpha^a_{\scriptscriptstyle NS} \cdot \alpha^a_{\scriptscriptstyle NL(l)}.
\label{ResponseDC-A-II}
\end{eqnarray}

This way, the detector's dual photon detection systems are like two disjoint detectors with a common receptacle.
It is by design that the {\it main system} holds the main calorimetry estimator power, whose ultimate precision is being optimised.
Instead, the {\it auxiliary system} collects less light over the same event samples.

Because both systems detect the same event, their calorimetry responses share some common, hence correlated,
systematics including $\alpha_{\scriptscriptstyle NL(l)}, \alpha_{\scriptscriptstyle NU}, \alpha_{\scriptscriptstyle NS}$. 
Instead, the main-to-auxiliary instrumental systematics $\alpha_{\scriptscriptstyle NL(q)}$ are designed to be different, i.e. as uncorrelated as they can possibly be.
Hence, the auxiliary \PCa\ system's uncorrelated and robust information is used to enhance the calorimetry systematics control of the main system.

In \DC, the common terms across the calorimetry responses of the $R^m$ and  $R^a$ can be cancelled across a ratio by construction or upon optimisation if needed.
Specifically, when comparing $R^m$ and  $R^a$, 
\begin{itemize}
\item
all common light generation effects (i.e. the $\alpha_{\scriptscriptstyle NL(l)}$ term) cancel, such as scintillation and Cherenkov radiation, so that $\alpha^m_{\scriptscriptstyle NL(l)} / \alpha^a_{\scriptscriptstyle NL(l)} \to 1$.
\item
the position-dependent effects (i.e. the $\alpha_{\scriptscriptstyle NU}$ term) are fixed to a constant given a static event vertex; and even though this constant varies with event vertex position, it can be calibrated, so 
$\alpha^m_{\scriptscriptstyle NU} / \alpha^a_{\scriptscriptstyle NU} \to 1$.
\item
the commonality on $\alpha_{\scriptscriptstyle NS}$ is due to the simultaneous detection of the same events in the common detector medium, and frequent calibration could monitor the variation of detector response over time so that
$\alpha^m_{\scriptscriptstyle NS} / \alpha^a_{\scriptscriptstyle NS} \to 1$.
\end{itemize}
Considering all the above premises, the \DC\ ratio is defined as
\begin{eqnarray}
\dfrac{ R^{m} }{ R^{a} } & = & 
	\dfrac{R^m_{\scriptscriptstyle o}}{R^a_{\scriptscriptstyle o}} 
	\dfrac{ \alpha^m_{\scriptscriptstyle NU} \cdot  \alpha^m_{\scriptscriptstyle NS} \cdot [\alpha^m_{\scriptscriptstyle NL(l)} \otimes \alpha^m_{\scriptscriptstyle NL(q)}] }{ \alpha^a_{\scriptscriptstyle NU} \cdot  \alpha^a_{\scriptscriptstyle NS} \cdot [\alpha^a_{\scriptscriptstyle NL(l)}] }.
\label{DCRatioGeneral}
\end{eqnarray}
By cancelling the correlated terms, i.e. $\alpha^m_{\scriptscriptstyle NU} / \alpha^a_{\scriptscriptstyle NU} \to 1$, $\alpha^m_{\scriptscriptstyle NS} / \alpha^a_{\scriptscriptstyle NS} \to 1$ and $\alpha^m_{\scriptscriptstyle NL(l)} / \alpha^a_{\scriptscriptstyle NL(l)} \to 1$, the \DC\ ratio transforms into
\begin{eqnarray}
\dfrac{ R^{m} }{ R^{a} } & = & 
	\dfrac{R^m_{\scriptscriptstyle o}}{R^a_{\scriptscriptstyle o}} 
	\alpha^m_{\scriptscriptstyle NL(q)},
\label{DCRatioQNL}
\end{eqnarray}
where $R^m_{\scriptscriptstyle o}$ and $R^a_{\scriptscriptstyle o}$ are constant to be calibrated, thus isolating the behaviour of $\alpha^m_{\scriptscriptstyle NL(q)}$.

Figure~\ref{Fig3.a}\blue{.(a.)} demonstrates the cancellation of all common responses and isolation of the channel-wise $\alpha^m_{\scriptscriptstyle NL(q)}$. 
Expressed in terms of energy, this 
leads to
\begin{eqnarray}
\dfrac{ E^{m} }{ E^{a} } & = & \dfrac{ E^{m}_{\scriptscriptstyle o}}{ E^{a}_{\scriptscriptstyle_o} }
	\alpha^m_{\scriptscriptstyle NL(q)} = \alpha^m_{\scriptscriptstyle NL(q)},
\label{DCRatioEQNL}
\end{eqnarray}
where $E^m_{\scriptscriptstyle o}$ and $E^a_{\scriptscriptstyle o}$ are the absolute anchored energy which are identical. 

Thus, the \DC\ ratio provides a clean and unbiased in-situ measurement of the so far elusive $\alpha^m_{\scriptscriptstyle NL(q)}$ term.
This term has been proven one of the most challenging and dominant contributions to the latest best precision experiments where their control of energy scale was controlled close to the $\sim$0.5\% precision~\cite{DYBcalib, DoubleChoozCalib}.

Moreover, the \DC\ opens a new handle for optimising the detector photon detection system, i.e. to maximise some $\alpha_{i}$ commonalities across both systems, thus maximising cancellations over the parameter space and isolating the parameters of interest.

\begin{figure*}[t!]
	\vglue 0.2cm
	\centering
        \includegraphics[scale=0.28]{Figure_4.pdf}
	\caption{
	{\bf (a.) Calorimetry Response Entanglement Illustration.} 
	A hypothetical channel-wise NL(q) model is assumed based on Figure~\ref{Fig3.a}\blue{.(a.)}. 
	{\bf (a.I)} The channel-wise NL(q) convolves into event-wise energy non-linearity with the events uniformly distributed in the full detection volume. 
	{\bf (a.II)} The position dependence of NL(q) makes it consequently entangling with NU, manifesting as a fake non-uniformity over detection volume, illustrated here with the events of 5 MeV energy. 
	{\bf (a.III)} The NL(q) could engender energy resolution deterioration due to biases in charge distribution and NU entanglement. The above biasing effects are more significant for $R^{m\scriptscriptstyle (\phi=20)}$ and also sizeable for $R^{m\scriptscriptstyle (\phi=8)}$. While in $R^{a\scriptscriptstyle (\phi=3)}$ or $R^{a\scriptscriptstyle (\phi=1)}$ configurations, the response orthogonality can be ensured hence negligible response entanglement biasing effects.  
	{\bf (b.) Calorimetry Response Control through Dual-Calorimetry.} 
    In $C^{m(\phi=20)}_{a(\phi=3 \ or \ 1)}$ and $C^{m(\phi=8)}_{a(\phi=3 \ or \ 1)}$  \DC\ configurations, after the channel-wise calibration for NL(q) as demonstrated in Figure~\ref{Fig3.b}\blue{.(b.)}, the NL(q) induced biases in energy NL, NU and resolution in \PIa\ calorimetry (i.e. $R^{m\scriptscriptstyle (\phi=20)}$ and $R^{m\scriptscriptstyle (\phi=8)}$) can be restricted within permille level, shown in {\bf (b.I) }, {\bf (b.II) } and {\bf (b.III)} respectively. 
	}	
	\label{Fig4}
\end{figure*}

\subsection*{ Dual Calorimetry Demonstration} 

The orchestration of the \DC\ with calibration sources and systems can be powerful since this allows to exploit high activity and high luminosity calibration sources such as laser or LED light sources, for accumulating high event statistics covering the entire photon dynamic range of interest within short periods.
By design, the calibration sources can be deployed at a given time and favourably in the central detector region,
so they may exploit the cancellations $\alpha^m_{\scriptscriptstyle NU} / \alpha^a_{\scriptscriptstyle NU} \to 1$ and $\alpha^m_{\scriptscriptstyle NS} / \alpha^a_{\scriptscriptstyle NS} \to 1$.
The scenario of a coherent \DC-based calibration~\cite{Han2020} offers the best context for the demonstration.

To quantitatively demonstrate the \DC, a Monte Carlo simulation-based analysis is performed with a hypothetical \LSDa\ detector design of the largest scale so far ($\sim2\times10^4$ m$^3$ volume), which can be considered as JUNO-like, as shown in Figure~\ref{Fig1.a}\blue{.(a.)}.
The typical energy of interest of LSD in neutrino physics, i.e. 1$\sim$10 MeV, is targeted.
This exercise aims to reach the energy scale control within permille level (up to the minimal 0.1\%) deviation over a vast photon ($n_{pd}$) dynamic range up to $O(10^{2})$, shown as ``LSD-XL'' in Figure~\ref{Fig2}, for combined response variations due to linearity and uniformity.

To render the exercise more representative, the {\it main photon detection system} will rely on SPS of 20-inch diameter in \PIa, which corresponds to the largest possible photo-multiplier tube (PMT) available in the market~\cite{Hama, WANG2012113} and used by most of the largest experiments now.
Its response is referred to as $R^{m\scriptscriptstyle (\phi=20)}$.
Similarly, the response of 8-inch SPS ($R^{m\scriptscriptstyle (\phi=8)}$) is also considered and leads to a reduced dynamic range response (i.e. $R^{m\scriptscriptstyle (\phi=8)}$).
As the {\it auxiliary photon detection system}, two SPS of diameters 3-inch and 1-inch are considered to increasingly match the \PCa\ condition, as indicated in Figure~\ref{Fig2}, whose respective responses will be noted as $R^{a\scriptscriptstyle (\phi=3)}$ and $R^{a\scriptscriptstyle (\phi=1)}$.

Hence, we shall consider four \DC\ configurations:
\begin{enumerate}

\item[a.]
    {\bf Configuration} $\mathbf{C^{m(\phi=20)}_{a(\phi=3)}}$,
    which combines the main system $R^{m\scriptscriptstyle (\phi=20)}$ of 15,000 20-inch SPS channels and the auxiliary system $R^{a\scriptscriptstyle (\phi=3)}$ of 50,000 3-inch SPS channels.

\item [b.]
    {\bf Configuration} $\mathbf{C^{m(\phi=8)}_{a(\phi=3)}}$,
    which combines the main system $R^{m\scriptscriptstyle (\phi=8)}$ of 100,000 8-inch SPS channels and the auxiliary system $R^{a\scriptscriptstyle (\phi=3)}$ of 50,000 3-inch SPS channels.

\item [c.]
    {\bf Configuration} $\mathbf{C^{m(\phi=20)}_{a(\phi=1)}}$,
    which combines the main system $R^{m\scriptscriptstyle (\phi=20)}$ of 15,000 20-inch SPS channels and the auxiliary system $R^{a\scriptscriptstyle (\phi=1)}$ of 100,000 1-inch SPS channels.

\item [d.]
    {\bf Configuration} $\mathbf{C^{m(\phi=8)}_{a(\phi=1)}}$,
    which combines the main system $R^{m\scriptscriptstyle (\phi=8)}$ of 100,000 8-inch SPS channels and the auxiliary system $R^{a\scriptscriptstyle (\phi=1)}$ of 100,000 1-inch SPS channels.

\end{enumerate}
where the total light level of
$R^{m\scriptscriptstyle (\phi=20)}$ and
$R^{m\scriptscriptstyle (\phi=8)}$ 
are both 1500 $n_{pd}$ per MeV, and that of $R^{a\scriptscriptstyle (\phi=3)}$ and $R^{a\scriptscriptstyle (\phi=1)}$ 
are, respectively,
100 and 
25 $n_{pd}$ per MeV.

\begin{figure*}[b!]
	\vglue 0.2cm
	\centering
	\includegraphics[scale=0.28]{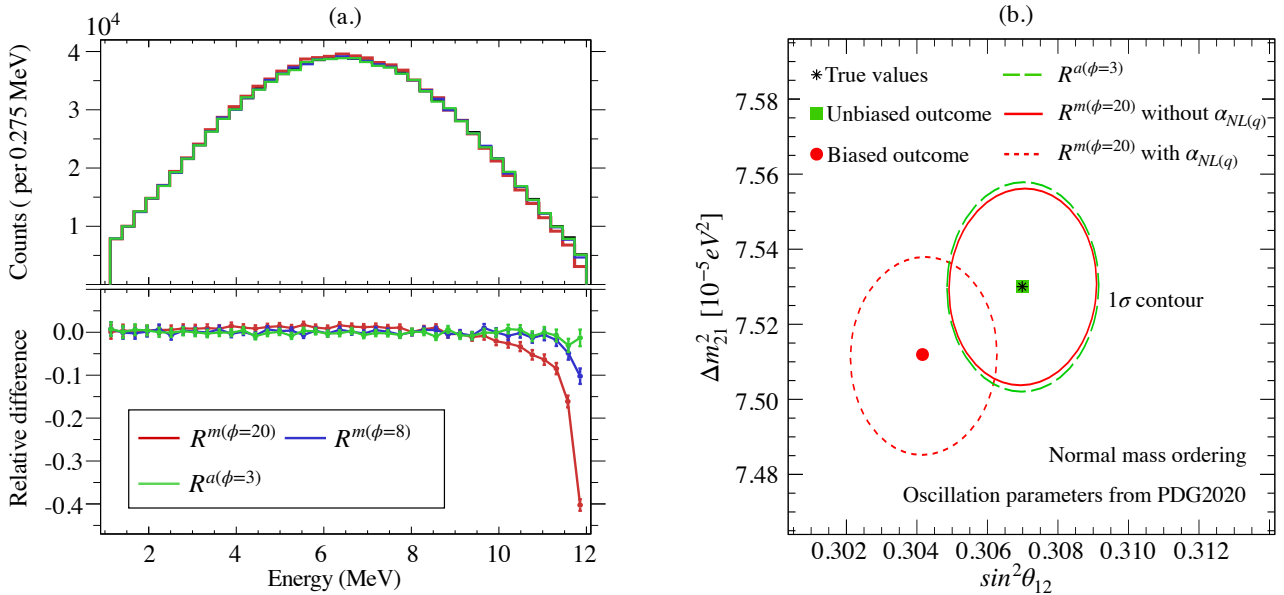}
	\caption
	{
	{\bf (a.) Distribution-wise Dual-Calorimetry Demonstration with $^{12}$B Spectrum.}
	Under \DC\ configuration, the cosmogenic-generated $^{12}B$ spectrum can be measured by the \PIa\ calorimetry indicated by the red and blue histograms in the top panel for $R^{m\scriptscriptstyle (\phi=20)}$  and $R^{m\scriptscriptstyle (\phi=8)}$ respectively; it can also be measured with the \PCa\ calorimetry indicated by the green histogram for $R^{a\scriptscriptstyle (\phi=3)}$ (or  $R^{a\scriptscriptstyle (\phi=1)}$ ).
 The impact of NL(q) introduced in Figure~\ref{Fig3.a}\blue{.(a.)} distorts the measured $^{12}B$ spectrum when using PIR, as opposed to the robust PCR
measurement, which can serve as a reference spectrum. 
	The measured spectrum is compared with the true spectrum, with the relative difference shown in the bottom panel. 
    {{\bf (b.) Dual-Calorimetry with Oscillation Parameters $\theta_{12}$ and $\Delta m^{2}_{21}$.}
   For the \DC\ in a JUNO-like experimental configuration~\cite{JUNO2021}, the oscillation parameters $\theta_{12}$ and $\Delta m^{2}_{21}$ can be measured through reactor neutrinos by both PIR calorimetry and PCR calorimetry.
   Under $C^{m(\phi=20)}_{a(\phi=3)}$ configuration, the $R^{m\scriptscriptstyle (\phi=20)}$ system has $\sim$3$\%/\sqrt{E}$ resolution, while $R^{a\scriptscriptstyle (\phi=3)}$ has $\sim$12$\%/\sqrt{E}$ resolution.
   Both systems can extract $\theta_{12}$ and $\Delta m^{2}_{21}$ with comparable precision (indicated by the 1$\sigma$ contours) from exactly the same data sample which, in this case, contains $\sim$100,000  events.
   For $R^{a\scriptscriptstyle (\phi=3)}$ with robust linear response, the extracted physics parameters are unbiased (green point) in comparison with the true values (black point).
   For $R^{m\scriptscriptstyle (\phi=20)}$, the introduced bias in energy as described in Figure~\ref{Fig4}\blue{.(a.I)} will bias the extracted physics parameters, as indicated by the red point.
   Therefore, the NL(q) induced bias in $R^{m\scriptscriptstyle (\phi=20)}$ can be diagnosed given the robust $\theta_{12}$ and $\Delta m^{2}_{21}$ measurement with $R^{a\scriptscriptstyle (\phi=3)}$.
   The statistic uncertainty and reactor-related uncertainties~\cite{JUNOprecision} are considered for this analysis. 
	}	
	}
	\label{Fig5}
\end{figure*}

For each configuration, a channel-wise calibration campaign is simulated with the following steps:

1.) A tunable light source (laser or led) is simulated at the detector centre with the intensity varying in light level from MeV to GeV equivalent. 
Each SPS of $R^{m}$ can be equally illuminated, covering its full dynamic range of interest, while each SPS of $R^{a}$ remains in PCR. 
Radioactive sources may also help investigate the delicate NL(q) effect due to particle-dependent liquid scintillator time properties \cite{LSDYB1}.

2.) By comparing the responses of $R^{m}$ and $R^{a}$, the $\alpha^m_{\scriptscriptstyle NL(q)}$ can be extracted for each channel with the help of the linear reference response of $R^{a}$ in PCR (i.e. $\alpha^a_{\scriptscriptstyle NL(q)} \rightarrow 1$), corresponding to Equation~(\ref{DCRatioGeneral}) and~(\ref{DCRatioQNL}), also as shown in Figure~\ref{Fig3.b}\blue{.(a.)}

3.) Given the extracted $\alpha^m_{\scriptscriptstyle NL(q)}$, corrections can be directly implemented on each channel and/or be performed through tuning the charge reconstruction of $R^{m}$ if allowed. \\

The above steps are illustrated in Figure~\ref{Fig3.b}, referred to as {\it channel-wise \DC\ calibration}, and the considered numerical input deviation $\alpha^m_{\scriptscriptstyle NL(q)}$ is hypothetical for academic illustration and conservative to possibly cover various kinds of realistic $\alpha^m_{\scriptscriptstyle NL(q)}$ forms.   
Given that, the potential of the dual calorimetry calibration can be quantified by reducing the original channel-wise $\alpha^m_{\scriptscriptstyle NL(q)}$ by one order of magnitude to permille level.

Furthermore, in terms of event-wise energy, as shown in Figure~\ref{Fig4}\blue{.(a.I)}, in parity of conditions, the $R^{m\scriptscriptstyle (\phi=20)}$ is more severely affected by $\alpha^m_{\scriptscriptstyle NL(q)}$ than the $R^{m\scriptscriptstyle (\phi=8)}$.
For events uniformly distributed in full volume, $R^{m\scriptscriptstyle (\phi=20)}$ undergoes $\sim$1\% event-wise NL(q) effect, while $\sim$0.3\% for $R^{m\scriptscriptstyle (\phi=8)}$.
For a typical energy of 5 MeV, the native $R^{m\scriptscriptstyle (\phi=20)}$ and $R^{m\scriptscriptstyle (\phi=8)}$ energy responses would suffer respectively $\sim$2\% and $\sim$0.5\% fake energy non-uniformity induced by NL(q) before the \DC\ calibration.
Those may be out of control within the \SC\ framework due to limited calibration sources and deployments.

With dual calorimetry calibration, the channel-wise $\alpha^m_{\scriptscriptstyle NL(q)}$ is disentangled and hence corrected, leading to $0.1\%$ control in terms of event-wise energy, as shown in Figure~\ref{Fig4}\blue{.(b.I)}; therefore the event-wise physics non-linearity $\alpha^m_{\scriptscriptstyle NL(l)}$ can be measured independently, e.g. in the detector centre.   
Once $\alpha^m_{\scriptscriptstyle NL}$ is fully known and therefore corrected, it ensures that the $\alpha^m_{\scriptscriptstyle NU}$ is extracted without any NL faked bias due to unaccounted correlations, as shown in Figure~\ref{Fig4}\blue{.(b.II)}. 
Thus, this channel-wise \DC\ exercise demonstrates how a more ``orthogonal'' calorimetric response basis (see Equation~(\ref{OrthogonalSC})) can be derived and regained in the context of a more stringent energy control precision without having to account for unknown correlations.


In addition, as shown in Figure~\ref{Fig4}\blue{.(a.III)}, the response bias due to $\alpha^m_{\scriptscriptstyle NL(q)}$ could lead to a sizeable energy resolution deterioration.
Again, this can be corrected by potentially up to one order of magnitude as shown in Figure~\ref{Fig4}\blue{.(b.III)}.
This performance is impressive if one considers how much extra light would be needed to compensate for an equivalent lost in resolution deterioration stochastically.
However, this is a regime historically never attained by \LSDa s.

Beyond the channel-wise approach, the \DC\ also works distribution-wise, i.e. the combination of all channels over numerous events.
This is illustrated, for instance, by using cosmogenic generated signals, such as the beta decay spectrum of $^{12}$B isotope, typically generated upon cosmic-muon spallation on $^{12}$C readily found in \LSDa s.
The main and auxiliary systems provide independent estimations on the same event sampling across the full detector volume and energy range of interest.
Hence, the distribution-wise approach enables the extraction of the overall $\alpha^m_{\scriptscriptstyle NL(q)}$ information, as illustrated in Figure~\ref{Fig5}\blue{.(a.)}.
This grants quantitative insight into the residual effects of the channel-wise approach after calibration for its persistent monitoring.

{The advantages of the \DC\ can go beyond the calibration context, as it may enable a unique framework for testing the energy response impact of the measurement of physics parameters.
This implies that the systematics of physics parameters can be diagnosed and quantified with extreme sensitivity relative to subtle detector response biases.
To illustrate this point, Figure~\ref{Fig5}\blue{.(b.)} shows the potential accuracy of the measurement of neutrino oscillation parameters $\theta_{12}$ and $\Delta m^2_{12}$, as obtained by \DC\ systems in a JUNO-like experimental configuration.}


Yet another advantage of the \DC\ design is that it enables an effective extension over the dynamic range of the overall detector.
The use of smaller SPS by the auxiliary photonics system allows the measurement of light at higher energies even when the main photonics system is limited by typical readout saturation effects. 
This feature is exploited in JUNO for the handling of cosmic-muons~\cite{Genster2018}, tagging the generation of cosmogenic isotopes (i.e. background), and even proton decay discovery searches~\cite{JUNOprotondecay}.

\section*{The Optimal Calorimetry Strategy} 

The presented multi-calorimetry detector discussion -- {\it single versus dual calorimetry} -- exploiting the advantages of different photon detection systems working in \PCa, \PIa\ or both, raises the question of whether there is an optimal design to yield the highest possible calorimetry precision. 

A single strict PCR calorimetry, benefiting from the immunity of charge measurement effects, hence minimal response entanglements, would be an attractive optimal candidate in terms of energy control.
In this context, the energy estimator is essentially digital rather than analogue, leading to minimal systematics.

The full digital readout may have non-negligible challenges at higher energies due to higher channel multiplicity.
The expected non-linearity due to Poissonian multiplicity saturation can be corrected, providing that not all channels are hit at once~\cite{DoubleChoozCalib}.
Such saturation effects may be further mitigated by reducing the photo-sensor dimensions, shortening the digitisation time unit or adding charge digitisation capability as redundancy.

Such a digital \SC\ detector also holds several other ideal features worth mentioning.
Smaller photo-sensors are known to benefit from much better resolution for vertex reconstruction resolution due to their better timing and effectively negligible dimensions.
Moreover, smaller photo-sensors are implosion resilient, as opposed to larger photo-sensors~\cite{DIWAN2012, LING2013}, thus reducing design overheads. 

However, for a realistic experiment, various factors including the cost in particular can lead to non-negligible compromises for calorimetry design.
In fact, the \DC\ can be regarded as one such compromise, where it maintains, as a priority, the dominant cost-effectiveness of the \PIa\ design, and achieves an effective optimal \PCa -based detector, given that the \PIa\ can be calibrated with \PCa\ as demonstrated above. 

Considering the optimal calorimetry detector design, it may indirectly address another interesting question, that is, whether the \DC\ concept could be further generalised towards more than two-photon detection systems.
Generally, increasing {\it multi-calorimetry} beyond the \DC\ concept cannot be ruled out a priori.
It has become more common for neutrino detectors to turn into effective multi-physics observatories, simultaneously addressing fundamental particle physics, astrophysics, and beyond.
This implies that one detector simultaneously targets different energy ranges with high precision.
Even though the condition behind the dual calorimetry can be fulfilled for a specific energy range, it can be violated for a different range. 
Hence, an additional photon detection system may help reach the PCR condition.
However, this discussion cannot be easily generalised and is specific to each 
experiment's conditions.

\section*{Conclusions} 

Much of the history of neutrino detection and physics is linked to advances in photonics technology and photon detection techniques for controlling the ultimate precision of observables, where energy measurement is most critical.
Most neutrino experiments so far have relied on \SC\ designs, even if they were not called so.
For present (e.g. JUNO) and future \LSDa s, the extreme control in energy pushes the traditional \SC\ boundaries beyond.
The \DC\ design, whose rationale is hereby described in full for the first time, is a possible upgrade over the \SC\ design, targeting for high-precision energy detection systematics control over vast dynamic range.
The core rationale is to endow the detector calorimetry with multiple synergic photon detection systems, thus leading to the breakdown of calorimetric response correlations.
In this framework, the design of the calibration scheme is critical for the optimisation and full exploitation of the \DC\ technique, as demonstrated by the channel-wise \DC\ calibration through which permille level systematic control in energy can be achieved robustly. 
The future of advances in light-based neutrino detectors remains tightly bound to advances in photon detection, where the \DC\ design is a practical and optimal design candidate beyond the full \PCa\ calorimetry design for achieving a high-precision calorimetry.

\section*{Acknowledgement} 

Anatael Cabrera acknowledges the key contributions of Yosuke Abe (PhD thesis), under close supervision by Masaki Ishitsuka and Masahiro Kuze, in the original developments of the so-called ``$\alpha$\,calibration'' for the Double Chooz experiment in tight collaboration between the APC (Paris, France) and Tokyo Institute of Technology (Tokyo, Japan).
Yang Han acknowledges the support of the National Natural Science Foundation of China with Grant No.12205391 and the Natural Science Foundation of Guangdong Province with Grant No.2023A1515012045.
The authors thank Double Chooz and JUNO collaborations' reviewing mechanisms for feedback.

\bibliography{biblio.bib}

\providecommand{\href}[2]{#2}\begingroup\raggedright\begin{thebibliography}{10}

\bibitem{Nobel1995}
F.~Reines and M.L.~Perl, \emph{The nobel prize in physics 1995},
  {\emph{www.nobelprize.org/prizes/ physics/1995/summary} }.

\bibitem{Nobel2002}
R.~Davis.Jr. and M.~Koshiba, \emph{The nobel prize in physics 2002.},
  {\emph{www.nobelprize.org/prizes/ physics/2002/summary} }.

\bibitem{Nobel2015}
T.~Kajita and A.B.~McDonald, \emph{The nobel prize in physics 2015},
  {\emph{www.nobelprize.org/prizes/ physics/2015/summary} }.

\bibitem{Birks}
J.B.~Birks, \emph{The Theory and Practice of Scintillation Counting}, Elsevier
  (1964),
  \href{https://doi.org/10.1016/c2013-0-01791-4}{10.1016/c2013-0-01791-4}.

\bibitem{Cherenkov1934}
P.A.~Cherenkov, \emph{{Visible luminescence of pure liquids under the influence
  of \ensuremath{\gamma}-radiation}},
  \href{https://doi.org/10.3367/UFNr.0093.196710n.0385}{\emph{Dokl. Akad. Nauk
  SSSR} {\bfseries 2} (1934) 451}.

\bibitem{Cherenkov1937}
P.A.~Cerenkov, \emph{{Visible radiation produced by electrons moving in a
  medium with velocities exceeding that of light}},
  \href{https://doi.org/10.1103/PhysRev.52.378}{\emph{Phys. Rev.} {\bfseries
  52} (1937) 378}.

\bibitem{JUNO2021}
{\scshape JUNO} collaboration, \emph{{JUNO physics and detector}},
  \href{https://doi.org/10.1016/j.ppnp.2021.103927}{\emph{Prog. Part. Nucl.
  Phys.} {\bfseries 123} (2022) 103927}
  [\href{https://arxiv.org/abs/arXiv:2104.02565}{{\ttfamily
  arXiv:2104.02565}}].

\bibitem{Cowan1956}
C.L.~Cowan et~al., \emph{{Detection of the free neutrino: A Confirmation}},
  \href{https://doi.org/10.1126/science.124.3212.103}{\emph{Science} {\bfseries
  124} (1956) 103}.

\bibitem{CHOOZ1999}
{\scshape CHOOZ} collaboration, \emph{{Limits on neutrino oscillations from the
  CHOOZ experiment}},
  \href{https://doi.org/10.1016/S0370-2693(99)01072-2}{\emph{Phys. Lett. B}
  {\bfseries 466} (1999) 415}
  [\href{https://arxiv.org/abs/arXiv:9907037}{{\ttfamily arXiv:9907037}}].

\bibitem{PaloVerde1999}
F.~Boehm et~al., \emph{{Search for neutrino oscillations at the Palo Verde
  nuclear reactors}},
  \href{https://doi.org/10.1103/PhysRevLett.84.3764}{\emph{Phys. Rev. Lett.}
  {\bfseries 84} (2000) 3764}
  [\href{https://arxiv.org/abs/arXiv:9912050}{{\ttfamily arXiv:9912050}}].

\bibitem{KamLAND2002}
{\scshape KamLAND} collaboration, \emph{{First results from KamLAND: Evidence
  for reactor anti-neutrino disappearance}},
  \href{https://doi.org/10.1103/PhysRevLett.90.021802}{\emph{Phys. Rev. Lett.}
  {\bfseries 90} (2003) 021802}
  [\href{https://arxiv.org/abs/arXiv:0212021}{{\ttfamily arXiv:0212021}}].

\bibitem{LSND1996}
{\scshape LSND} collaboration, \emph{{Evidence for anti-muon-neutrino to
  anti-electron-neutrino oscillations from the LSND experiment at LAMPF}},
  \href{https://doi.org/10.1103/PhysRevLett.77.3082}{\emph{Phys. Rev. Lett.}
  {\bfseries 77} (1996) 3082}
  [\href{https://arxiv.org/abs/arXiv:9605003}{{\ttfamily arXiv:9605003}}].

\bibitem{KARMEN2002}
{\scshape KARMEN} collaboration, \emph{{Upper limits for neutrino oscillations
  muon-anti-neutrino to electron-anti-neutrino from muon decay at rest}},
  \href{https://doi.org/10.1103/PhysRevD.65.112001}{\emph{Phys. Rev. D}
  {\bfseries 65} (2002) 112001}
  [\href{https://arxiv.org/abs/arXiv:0203021}{{\ttfamily arXiv:0203021}}].

\bibitem{KamLANDgeo}
T.~Araki et~al., \emph{{Experimental investigation of geologically produced
  antineutrinos with KamLAND}},
  \href{https://doi.org/10.1038/nature03980}{\emph{Nature} {\bfseries 436}
  (2005) 499}.

\bibitem{DayaBay2012}
{\scshape Daya Bay} collaboration, \emph{{Observation of electron-antineutrino
  disappearance at Daya Bay}},
  \href{https://doi.org/10.1103/PhysRevLett.108.171803}{\emph{Phys. Rev. Lett.}
  {\bfseries 108} (2012) 171803}
  [\href{https://arxiv.org/abs/arXiv:1203.1669}{{\ttfamily arXiv:1203.1669}}].

\bibitem{DoubleChooz2011}
{\scshape Double Chooz} collaboration, \emph{{Indication of Reactor
  $\bar{\nu}_e$ Disappearance in the Double Chooz Experiment}},
  \href{https://doi.org/10.1103/PhysRevLett.108.131801}{\emph{Phys. Rev. Lett.}
  {\bfseries 108} (2012) 131801}
  [\href{https://arxiv.org/abs/arXiv:1112.6353}{{\ttfamily arXiv:1112.6353}}].

\bibitem{RENO2012}
{\scshape RENO} collaboration, \emph{{Observation of Reactor Electron
  Antineutrino Disappearance in the RENO Experiment}},
  \href{https://doi.org/10.1103/PhysRevLett.108.191802}{\emph{Phys. Rev. Lett.}
  {\bfseries 108} (2012) 191802}
  [\href{https://arxiv.org/abs/arXiv:1204.0626}{{\ttfamily arXiv:1204.0626}}].

\bibitem{BorexinoBe7}
{\scshape Borexino} collaboration, \emph{{Direct Measurement of the Be-7 Solar
  Neutrino Flux with 192 Days of Borexino Data}},
  \href{https://doi.org/10.1103/PhysRevLett.101.091302}{\emph{Phys. Rev. Lett.}
  {\bfseries 101} (2008) 091302}
  [\href{https://arxiv.org/abs/arXiv:0805.3843}{{\ttfamily arXiv:0805.3843}}].

\bibitem{BorexinoB8}
{\scshape Borexino} collaboration, \emph{{Measurement of the solar 8B neutrino
  rate with a liquid scintillator target and 3 MeV energy threshold in the
  Borexino detector}},
  \href{https://doi.org/10.1103/PhysRevD.82.033006}{\emph{Phys. Rev. D}
  {\bfseries 82} (2010) 033006}
  [\href{https://arxiv.org/abs/arXiv:0808.2868}{{\ttfamily arXiv:0808.2868}}].

\bibitem{Borexinopep}
{\scshape Borexino} collaboration, \emph{{First evidence of pep solar neutrinos
  by direct detection in Borexino}},
  \href{https://doi.org/10.1103/PhysRevLett.108.051302}{\emph{Phys. Rev. Lett.}
  {\bfseries 108} (2012) 051302}
  [\href{https://arxiv.org/abs/arXiv:1110.3230}{{\ttfamily arXiv:1110.3230}}].

\bibitem{Borexinopp}
{\scshape Borexino} collaboration, \emph{{Comprehensive measurement of
  $pp$-chain solar neutrinos}},
  \href{https://doi.org/10.1038/s41586-018-0624-y}{\emph{Nature} {\bfseries
  562} (2018) 505}.

\bibitem{BorexinoCNO}
{\scshape Borexino} collaboration, \emph{{Experimental evidence of neutrinos
  produced in the CNO fusion cycle in the Sun}},
  \href{https://doi.org/10.1038/s41586-020-2934-0}{\emph{Nature} {\bfseries
  587} (2020) 577} [\href{https://arxiv.org/abs/arXiv:2006.15115}{{\ttfamily
  arXiv:2006.15115}}].

\bibitem{KamLANDZen2022}
{\scshape KamLAND-Zen} collaboration, \emph{{Search for the Majorana Nature of
  Neutrinos in the Inverted Mass Ordering Region with KamLAND-Zen}},
  \href{https://doi.org/10.1103/PhysRevLett.130.051801}{\emph{Phys. Rev. Lett.}
  {\bfseries 130} (2023) 051801}
  [\href{https://arxiv.org/abs/arXiv:2203.02139}{{\ttfamily
  arXiv:2203.02139}}].

\bibitem{SNOplus2021}
{\scshape SNO+} collaboration, \emph{{The SNO+ experiment}},
  \href{https://doi.org/10.1088/1748-0221/16/08/P08059}{\emph{JINST} {\bfseries
  16} (2021) P08059} [\href{https://arxiv.org/abs/arXiv:2104.11687}{{\ttfamily
  arXiv:2104.11687}}].

\bibitem{SuperK1998}
{\scshape Super-Kamiokande} collaboration, \emph{{Evidence for oscillation of
  atmospheric neutrinos}},
  \href{https://doi.org/10.1103/PhysRevLett.81.1562}{\emph{Phys. Rev. Lett.}
  {\bfseries 81} (1998) 1562}
  [\href{https://arxiv.org/abs/arXiv:9807003}{{\ttfamily arXiv:9807003}}].

\bibitem{SNO2002}
{\scshape SNO} collaboration, \emph{{Direct evidence for neutrino flavor
  transformation from neutral current interactions in the Sudbury Neutrino
  Observatory}},
  \href{https://doi.org/10.1103/PhysRevLett.89.011301}{\emph{Phys. Rev. Lett.}
  {\bfseries 89} (2002) 011301}
  [\href{https://arxiv.org/abs/arXiv:0204008}{{\ttfamily arXiv:0204008}}].

\bibitem{Kamiokande1987}
{\scshape Kamiokande-II} collaboration, \emph{{Observation of a Neutrino Burst
  from the Supernova SN 1987a}},
  \href{https://doi.org/10.1103/PhysRevLett.58.1490}{\emph{Phys. Rev. Lett.}
  {\bfseries 58} (1987) 1490}.

\bibitem{IMB1987}
R.M.~Bionta et~al., \emph{{Observation of a Neutrino Burst in Coincidence with
  Supernova SN 1987a in the Large Magellanic Cloud}},
  \href{https://doi.org/10.1103/PhysRevLett.58.1494}{\emph{Phys. Rev. Lett.}
  {\bfseries 58} (1987) 1494}.

\bibitem{HyperK2018}
{\scshape Hyper-Kamiokande} collaboration, \emph{{Hyper-Kamiokande Design
  Report}},  \href{https://arxiv.org/abs/[arXiv:1805.04163]}{{\ttfamily
  [arXiv:1805.04163]}}.

\bibitem{IceCubeDesign}
{\scshape IceCube} collaboration, \emph{{The IceCube Neutrino Observatory:
  Instrumentation and Online Systems}},
  \href{https://doi.org/10.1088/1748-0221/12/03/P03012}{\emph{JINST} {\bfseries
  12} (2017) P03012} [\href{https://arxiv.org/abs/arXiv:1612.05093}{{\ttfamily
  arXiv:1612.05093}}].

\bibitem{KM3Net2016}
{\scshape KM3Net} collaboration, \emph{{Letter of intent for KM3NeT 2.0}},
  \href{https://doi.org/10.1088/0954-3899/43/8/084001}{\emph{J. Phys. G}
  {\bfseries 43} (2016) 084001}
  [\href{https://arxiv.org/abs/arXiv:1601.07459}{{\ttfamily
  arXiv:1601.07459}}].

\bibitem{Baikal}
B.~Collaboration, \emph{{Deep-Underwater Cherenkov Detector in Lake Baikal}},
  \href{https://doi.org/10.1134/S1063776122040148}{\emph{J. Exp. Theor. Phys.}
  {\bfseries 134} (2022) 399}.

\bibitem{JUNO_YB}
F.~An and et~al., \emph{Neutrino physics with juno},
  \href{https://doi.org/10.1088/0954-3899/43/3/030401}{\emph{Journal of Physics
  G: Nuclear and Particle Physics} {\bfseries 43} (2016) 030401}
  [\href{https://arxiv.org/abs/arXiv:1507.05613}{{\ttfamily
  arXiv:1507.05613}}].

\bibitem{Anatael2016}
A.~Cabrera, \emph{High precision calorimetry with liquid scintillator
  detectors}, \href{https://doi.org/10.5281/zenodo.8268175}{\emph{Talk at
  Frontiers of liquid Scintillator Technology workshop, 10.5281/zenodo.8268175}
  (2016) }.

\bibitem{Anatael2018}
A.~Cabrera, \emph{Juno stereo-calorimetry},
  \href{https://doi.org/10.5281/zenodo.1314425}{\emph{Talk at Energy Scale
  Calibration in Anti-neutrino Precision Experiments workshop,
  10.5281/zenodo.1314425} (2018) }.

\bibitem{CUPID2019}
{\scshape CUPID} collaboration, \emph{{CUPID pre-CDR}},
  \href{https://arxiv.org/abs/[arXiv:1907.09376]}{{\ttfamily
  [arXiv:1907.09376]}}.

\bibitem{nEXO}
{\scshape nEXO} collaboration, \emph{{nEXO Pre-Conceptual Design Report}},
  \href{https://arxiv.org/abs/[arXiv:1805.11142]}{{\ttfamily
  [arXiv:1805.11142]}}.

\bibitem{BorexinoPSD}
{\scshape Borexino Collaboration} collaboration, \emph{Simultaneous precision
  spectroscopy of $pp$, $^{7}\mathrm{Be}$, and $pep$ solar neutrinos with
  borexino phase-ii},
  \href{https://doi.org/10.1103/PhysRevD.100.082004}{\emph{Phys. Rev. D}
  {\bfseries 100} (2019) 082004}
  [\href{https://arxiv.org/abs/arXiv:1707.09279}{{\ttfamily
  arXiv:1707.09279}}].

\bibitem{LiquidO2019}
{\scshape LiquidO} collaboration, \emph{{Neutrino Physics with an Opaque
  Detector}}, \href{https://doi.org/10.1038/s42005-021-00763-5}{\emph{Commun.
  Phys.} {\bfseries 4} (2021) 273}
  [\href{https://arxiv.org/abs/arXiv:1908.02859}{{\ttfamily
  arXiv:1908.02859}}].

\bibitem{NOvA2004}
{\scshape NOvA} collaboration, \emph{{NOvA: Proposal to Build a 30 Kiloton
  Off-Axis Detector to Study $\nu_{\mu} \to \nu_e$ Oscillations in the NuMI
  Beamline}},  \href{https://arxiv.org/abs/[arXiv:0503053]}{{\ttfamily
  [arXiv:0503053]}}.

\bibitem{MINOS_detector}
{\scshape MINOS} collaboration, \emph{{The Magnetized steel and scintillator
  calorimeters of the MINOS experiment}},
  \href{https://doi.org/10.1016/j.nima.2008.08.003}{\emph{Nucl. Instrum. Meth.
  A} {\bfseries 596} (2008) 190}
  [\href{https://arxiv.org/abs/arXiv:0805.3170}{{\ttfamily arXiv:0805.3170}}].

\bibitem{T2Knear}
{\scshape T2K} collaboration, \emph{{T2K ND280 Upgrade - Technical Design
  Report}},  \href{https://arxiv.org/abs/[arXiv:1901.03750]}{{\ttfamily
  [arXiv:1901.03750]}}.

\bibitem{SoLid}
{\scshape SoLid} collaboration, \emph{{SoLid: a short baseline reactor neutrino
  experiment}},
  \href{https://doi.org/10.1088/1748-0221/16/02/P02025}{\emph{JINST} {\bfseries
  16} (2021) P02025} [\href{https://arxiv.org/abs/arXiv:2002.05914}{{\ttfamily
  arXiv:2002.05914}}].

\bibitem{DANSS}
I.~Alekseev et~al., \emph{Search for sterile neutrinos at the danss
  experiment},
  \href{https://doi.org/https://doi.org/10.1016/j.physletb.2018.10.038}{\emph{Phys
  Lett. B} {\bfseries 787} (2018) 56}
  [\href{https://arxiv.org/abs/arXiv:1804.04046}{{\ttfamily
  arXiv:1804.04046}}].

\bibitem{Borexino_cherenkov}
{\scshape Borexino} collaboration, \emph{{Correlated and integrated
  directionality for sub-MeV solar neutrinos in Borexino}},
  \href{https://doi.org/10.1103/PhysRevD.105.052002}{\emph{Phys. Rev. D}
  {\bfseries 105} (2022) 052002}
  [\href{https://arxiv.org/abs/arXiv:2109.04770}{{\ttfamily
  arXiv:2109.04770}}].

\bibitem{SNO2022}
{\scshape SNO+} collaboration, \emph{{Evidence of Antineutrinos from Distant
  Reactors using Pure Water at SNO+}},
  \href{https://doi.org/10.1103/PhysRevLett.130.091801}{\emph{Phys. Rev. Lett.}
  {\bfseries 130} (2023) 091801}
  [\href{https://arxiv.org/abs/arXiv:2210.14154}{{\ttfamily
  arXiv:2210.14154}}].

\bibitem{WBLS}
J.R.~Alonso et~al., \emph{{Advanced Scintillator Detector Concept (ASDC): A
  Concept Paper on the Physics Potential of Water-Based Liquid Scintillator}},
  \href{https://arxiv.org/abs/[arXiv:1409.5864]}{{\ttfamily
  [arXiv:1409.5864]}}.

\bibitem{Yeh2011}
M.~Yeh et~al., \emph{{A new water-based liquid scintillator and potential
  applications}}, \href{https://doi.org/10.1016/j.nima.2011.08.040}{\emph{Nucl.
  Instrum. Meth. A} {\bfseries 660} (2011) 51}.

\bibitem{Theia}
{\scshape Theia} collaboration, \emph{{THEIA: an advanced optical neutrino
  detector}}, \href{https://doi.org/10.1140/epjc/s10052-020-7977-8}{\emph{Eur.
  Phys. J. C} {\bfseries 80} (2020) 416}
  [\href{https://arxiv.org/abs/arXiv:1911.03501}{{\ttfamily
  arXiv:1911.03501}}].

\bibitem{SlowFluor}
S.D.~Biller et~al., \emph{{Slow fluors for effective separation of Cherenkov
  light in liquid scintillators}},
  \href{https://doi.org/10.1016/j.nima.2020.164106}{\emph{Nucl. Instrum. Meth.
  A} {\bfseries 972} (2020) 164106}
  [\href{https://arxiv.org/abs/arXiv:2001.10825}{{\ttfamily
  arXiv:2001.10825}}].

\bibitem{SlowLS}
Z.~Guo et~al., \emph{{Slow Liquid Scintillator Candidates for MeV-scale
  Neutrino Experiments}},
  \href{https://doi.org/10.1016/j.astropartphys.2019.02.001}{\emph{Astropart.
  Phys.} {\bfseries 109} (2019) 33}
  [\href{https://arxiv.org/abs/arXiv:1708.07781}{{\ttfamily
  arXiv:1708.07781}}].

\bibitem{FADC}
{Wikipedia}, \emph{Flash adc --- {Wikipedia}{,} the free encyclopedia},  2023.

\bibitem{waveform_analog}
G.~Haller and B.~Wooley, \emph{An analog memory integrated circuit for waveform
  sampling up to 900 mhz}, \href{https://doi.org/10.1109/23.322884}{\emph{IEEE
  Transactions on Nuclear Science} {\bfseries 41} (1994) 1203}.

\bibitem{DYBcalib}
{\scshape Daya Bay} collaboration, \emph{{A high precision calibration of the
  nonlinear energy response at Daya Bay}},
  \href{https://doi.org/10.1016/j.nima.2019.06.031}{\emph{Nucl. Instrum. Meth.
  A} {\bfseries 940} (2019) 230}
  [\href{https://arxiv.org/abs/arXiv:1902.08241}{{\ttfamily
  arXiv:1902.08241}}].

\bibitem{DCNature}
{\scshape Double Chooz} collaboration, \emph{{Double Chooz $\theta_{13}$
  measurement via total neutron capture detection}},
  \href{https://doi.org/10.1038/s41567-020-0831-y}{\emph{Nature Phys.}
  {\bfseries 16} (2020) 558}
  [\href{https://arxiv.org/abs/arXiv:1901.09445}{{\ttfamily
  arXiv:1901.09445}}].

\bibitem{JUNOcalibration}
{\scshape JUNO} collaboration, \emph{{Calibration Strategy of the JUNO
  Experiment}}, \href{https://doi.org/10.1007/JHEP03(2021)004}{\emph{JHEP}
  {\bfseries 03} (2021) 004}
  [\href{https://arxiv.org/abs/arXiv:2011.06405}{{\ttfamily
  arXiv:2011.06405}}].

\bibitem{NL2015}
Z.~Fei-Hong et~al., \emph{Measurement of the liquid scintillator nonlinear
  energy response to electron},
  \href{https://doi.org/10.1088/1674-1137/39/1/016003}{\emph{Chin. Phys. C}
  {\bfseries 39} (2015) 016003}
  [\href{https://arxiv.org/abs/arXiv:1403.3257}{{\ttfamily arXiv:1403.3257}}].

\bibitem{DoubleChoozproposal}
F.~Ardellier et~al., \emph{{Letter of intent for Double-CHOOZ: A Search for the
  mixing angle theta(13)}},
  \href{https://arxiv.org/abs/[arXiv:0405032]}{{\ttfamily [arXiv:0405032]}}.

\bibitem{DayaBaydetector}
{\scshape Daya Bay} collaboration, \emph{{The Detector System of The Daya Bay
  Reactor Neutrino Experiment}},
  \href{https://doi.org/10.1016/j.nima.2015.11.144}{\emph{Nucl. Instrum. Meth.
  A} {\bfseries 811} (2016) 133}
  [\href{https://arxiv.org/abs/arXiv:1508.03943}{{\ttfamily
  arXiv:1508.03943}}].

\bibitem{RENOexperiment}
{\scshape RENO} collaboration, \emph{{RENO: An Experiment for Neutrino
  Oscillation Parameter $\theta_{13}$ Using Reactor Neutrinos at Yonggwang}},
  \href{https://arxiv.org/abs/[arXiv:1003.1391]}{{\ttfamily
  [arXiv:1003.1391]}}.

\bibitem{DoubleChoozCalib}
E.~Chauveau, \emph{Calibration and energy scale in double chooz},
  \href{https://doi.org/10.5281/zenodo.1314380}{\emph{Talk at Energy Scale
  Calibration in Anti-neutrino Precision Experiments workshop,
  10.5281/zenodo.1314380} (2018) }.

\bibitem{Han2020}
Y.~Han, \emph{{PhD Thesis: Dual Calorimetry for High Precision Neutrino
  Oscillation Measurement at JUNO Experiment}},
  \href{https://arxiv.org/abs/[HAL:tel-03295420]}{{\ttfamily
  [HAL:tel-03295420]}}.

\bibitem{Hama}
Hamamatsu, \emph{www.hamamatsu.com/us/en/why-hamamatsu/20inch-pmts}, .

\bibitem{WANG2012113}
Y.~Wang et~al., \emph{A new design of large area mcp-pmt for the next
  generation neutrino experiment},
  \href{https://doi.org/https://doi.org/10.1016/j.nima.2011.12.085}{\emph{Nucl.
  Instrum. Meth. A} {\bfseries 695} (2012) 113}.

\bibitem{JUNOprecision}
{\scshape JUNO} collaboration, \emph{{Sub-percent precision measurement of
  neutrino oscillation parameters with JUNO}},
  \href{https://doi.org/10.1088/1674-1137/ac8bc9}{\emph{Chin. Phys. C}
  {\bfseries 46} (2022) 123001}
  [\href{https://arxiv.org/abs/arXiv:2204.13249}{{\ttfamily
  arXiv:2204.13249}}].

\bibitem{LSDYB1}
X.~Li et~al., \emph{Timing properties and pulse shape discrimination of
  {LAB}-based liquid scintillator},
  \href{https://doi.org/10.1088/1674-1137/35/11/009}{\emph{Chin. Phys. C}
  {\bfseries 35} (2011) 1026}.

\bibitem{Genster2018}
C.~Genster et~al., \emph{{Muon reconstruction with a geometrical model in
  JUNO}}, \href{https://doi.org/10.1088/1748-0221/13/03/T03003}{\emph{JINST}
  {\bfseries 13} (2018) T03003}
  [\href{https://arxiv.org/abs/arXiv:1906.01912}{{\ttfamily
  arXiv:1906.01912}}].

\bibitem{JUNOprotondecay}
{\scshape JUNO} collaboration, \emph{{JUNO Sensitivity on Proton Decay $p\to
  \bar\nu K^+$ Searches}},
  \href{https://doi.org/10.1088/1674-1137/ace9c6}{\emph{Chin. Phys. C}
  {\bfseries 47} (2023) 113002}
  [\href{https://arxiv.org/abs/arXiv:2212.08502}{{\ttfamily
  arXiv:2212.08502}}].

\bibitem{DIWAN2012}
D.~Milind and et~al, \emph{Underwater implosions of large format
  photo-multiplier tubes},
  \href{https://doi.org/https://doi.org/10.1016/j.nima.2011.12.033}{\emph{Nucl.
  Instrum. Meth. A} {\bfseries 670} (2012) 61}.

\bibitem{LING2013}
J.~Ling. et~al., \emph{Implosion chain reaction mitigation in underwater
  assemblies of photomultiplier tubes},
  \href{https://doi.org/https://doi.org/10.1016/j.nima.2013.07.056}{\emph{Nucl.
  Instrum. Meth. A} {\bfseries 729} (2013) 491}.

\end{thebibliography}\endgroup

\end{multicols*}

\end{document}